\documentclass[journal,comsoc]{IEEEtran}
\usepackage[ruled,lined]{algorithm2e}
\usepackage{xcolor, soul}
\sethlcolor{yellow}
\usepackage{mathrsfs}  
\usepackage{graphicx}
\usepackage{xcolor}
\usepackage{graphicx,epsf,epic,eepic,psfig,latexcad}
\renewcommand{\baselinestretch}{1}

\usepackage{subcaption}
\usepackage{amssymb,amsmath,caption}
\usepackage{cite}
\usepackage{amsmath, amssymb, amsthm}
\usepackage{graphicx}
\renewcommand{\baselinestretch}{1}
\usepackage{epstopdf}
\DeclareMathOperator*{\argmin}{arg\,min} 
\makeatletter
\usepackage{textcomp}
\makeatother
\usepackage{xcolor}
\begin{document}
	
	\title{Random Spreading for Unsourced MAC with  Power Diversity}
	\author{Mohammad Javad Ahmadi and Tolga M. Duman \vspace{-6mm}
		\thanks{This research is funded by the Scientific and Technological Research Council of Turkey (TUBITAK) under the grant 119E589.}
		\thanks{The authors are with the Department of Electrical and Electronics Engineering, Bilkent University, 06800 Ankara, Turkey (e-mail: ahmadi@ee.bilkent.edu.tr; duman@ee.bilkent.edu.tr).}
	}
	
	\maketitle
	\begin{abstract}
		We propose an improvement of the random spreading approach with polar codes for unsourced multiple access, for which each user first encodes its message by a polar code, and then the coded bits are spread using a random spreading sequence. The proposed approach divides the active users into different groups, and employs different power levels for each group in such a way that the average power constraint is satisfied. We formulate and solve an optimization problem to determine the number of groups, and the corresponding user numbers and power levels. Extensive simulations show that the proposed approach outperforms the existing methods, especially when the number of active users is large.
	\end{abstract}
	\vspace{-1.8mm}
	\section{Introduction}
	Next generation wireless systems require effective solutions for massive random access motivating the setting of unsourced multiple access (MAC)~\cite{polyanskiy2017perspective}. In unsourced MAC, a large number of uncoordinated users share a common channel to convey their messages to a central unit. At any given time, only a small subset of all the potential users are active, and the messages are short. Since there is no coordination among the users and there are a very large number of them, there is no user identification, and the users employ the same codebook to convey their messages (hence the name {\em unsourced MAC}). The receiver is only interested in decoding the messages transmitted by the users without any regard to the user identities, and the per-user probability of error (PUPE) criterion is adopted as the performance metric. 
	
	\indent The formulation of unsourced MAC and its fundamental limits of communication are developed in \cite{polyanskiy2017perspective}. Several low-complexity schemes aiming at reaching the finite-blocklength achievability bounds in \cite{polyanskiy2017perspective} are devised in the literature. These include solutions based on T-fold Aloha \cite{calderbank2020chirrup, vem2017user, vem2019user, ordentlich2017low,marshakov2019polar}, splitting the payload \cite{amalladinne2018coupled,amalladinne2018coded,9432925}, sparsifying codeword bits \cite{pradhan2019joint,zheng2020polar,truhachev2020low, tanc2021massive, han2021sparse,ebert2021stochastic}, and random spreading \cite{pradhan2020polar, pradhan2021ldpc} approaches. 
	
	\indent Among these approaches, the best performance is obtained by the one based on the random spreading idea in \cite{pradhan2020polar}, particularly, for the number of active users $K_a \leq 225$. Unlike most papers that use short block length transmissions to deal with interference, \cite{pradhan2020polar} considers a scheme where users employ transmissions covering the entire frame. To confront the interference, each user's message is spread via random sequences. Although the scheme performs very well for the relatively low number of active users, its performance deteriorates significantly when $K_a$ exceeds a certain threshold. The reason for the decline in performance is that the interference is treated as noise and the increase in the number of interfering users beyond some threshold becomes excessive.
	
	\indent In another related work \cite{zheng2020polar}, a new polar coding scheme based on sparse spreading is proposed. The codewords are obtained by different codes of varying lengths, and they are modulated using different power levels. The authors experimentally determine the number of power levels to be used as well as the code lengths and the specific power levels. This scheme offers improvements over the solution in \cite{pradhan2020polar} for the larger number of users, specifically, it outperforms the existing unsourced MAC solutions for $250\leq K_a\leq  300$. 
	
	\indent In this letter, we propose an extension of the random spreading idea in \cite{pradhan2020polar} by allowing for the use of different power levels among active users. We formulate an optimization problem to determine the optimal number of power levels as well as their specific values. Each user independently determines which power level to use with a certain probability in such a way to match the optimal power level statistics. We demonstrate that interference is well mitigated by this approach, and the unsourced MAC performance becomes highly superior for a wide range of the number of active users. We also present a simple detector based on an estimate of the covariance matrix of the received signal as an alternative to the one in \cite{pradhan2020polar}. Specifically, the poor performance of the scheme in \cite{pradhan2020polar} is ameliorated for $K_a>225$. The proposed analytical approach can also be used for determining the power levels in the scheme of \cite{zheng2020polar}, eliminating the need for extensive numerical experiments for system design. Furthermore, the basic ideas and methodology can potentially be applied to other unsourced MAC schemes as well.

	\indent The letter is organized as follows. In Section II, we describe the system model. In Section III, the proposed random access scheme is described. In Section IV, an optimization problem for power allocation (PA) is formulated and solved. Simulation results are provided in Section V, and our conclusions are given in Section VI.
	\vspace{-2.8mm}
	\section{System Model}
	
	We consider an unsourced random access model in which $K_a$ out of $K_T$ users are active at a given frame. The number of active users $K_a$ is assumed to be known at the receiver. Each active user transmits $B$ bits of information through $n$ channel uses. Assuming a Gaussian multiple-access channel (GMAC), the received signal vector in the absence of synchronization errors and fading is written as
	\begin{align}
	\mathbf{y} = \sum_{i=1}^{K_T}{s_i\mathbf{x}_i(\mathbf{u}_i)+{\mathbf{z}}},
	\label{eqs1}
	\end{align}
	where $s_i$ is 1 for active users and $0$ otherwise, $\mathbf{x}_i$ is the length $n$ spread polar-coded signal  corresponding to the message bit sequence $\mathbf{u}_i\in \{0,1\}^B$ of user $i$, and ${\mathbf{z}}\sim \mathcal{N}(\mathbf{0}, \mathbf{I}_n)$ is the additive white Gaussian noise. Each user selects its message index uniformly from the set $\{i\in\mathbb{Z}:1\leq i\leq 2^B \}$. The average power of each user per channel use is set to $P$. Therefore, the energy-per-bit of the system can be written as
	\begin{align}
	\dfrac{E_b}{N_0}=\dfrac{nP}{2B},
	\end{align}
	\noindent and the PUPE of the system is defined as
	\begin{align}
	P_e = \max_{\sum_{i=1}^{K_T}{s_i}=K_a} \dfrac{1}{K_a}\sum_{i=1}^{K_T}{s_i \mathrm{Pr}(\mathbf{u}_i\notin \mathcal{L}(\mathbf{y}))},
	\end{align}
	\noindent where $\mathcal{L}(\mathbf{y})$ is the list of decoded messages with size at most $K_a$. The primary purpose is to design encoding and decoding schemes to reach a PUPE less than the target block error probability $\epsilon$ with the lowest ${E_b}/{N_0}$.
	\vspace{-2.8mm}
	\section{Proposed Unsourced MAC Scheme}
	\vspace{2.5mm}
	\subsection{Encoder}
	The message selected by each user is divided into two parts with $B_s$ and $B_c=B-B_s$ bits, i.e., $\mathbf{u}_i=(\mathbf{u}_{is},\mathbf{u}_{ic})$. The first part is used to map the preamble bits, $\mathbf{u}_{is}$, to columns of a signature codebook $\mathbf{A}\in \mathbb{R}^{n_s, 2^{B_s}}$, where $n_s$ is the spreading sequence length. Since
	\begin{align}
	P[ \exists j\neq i :\mathbf{u}_{is}=\mathbf{u}_{js}]\leq \dfrac{K_a -1}{2^{B_s}}, \label{eqss5}
	\end{align}
	$B_s$ is selected to make the right-hand side of \eqref{eqss5} in such a way to satisfy the PUPE requirement. The elements of the matrix $\mathbf{A}$ are generated by first picking independent zero-mean Gaussian random variables with unit variance, forcing each column to have an average of zero by subtracting its mean from each element, and then scaling them to distinct power levels (the determination of the power levels will be discussed in the following section). Let us denote the column picked by the $i$th user by $\mathbf{a}_i$. The second part of the message, $\mathbf{u}_{ic}$, is encoded using a polar code. As in \cite{pradhan2020polar}, this message is appended by $r$ cyclic redundancy check (CRC) bits, and the result is passed to an $(n_c, B_c+r)$ polar encoder. The CRC bits are used to check the success of polar decoding. The polar codeword is modulated using binary shift keying (BPSK), resulting in $\mathbf{v}_i\in \{\pm 1\}^{n_c\times 1}$. The transmitted signal for user $i$ is then obtained as 
	\begin{align}
	\mathbf{x}_i = \mathbf{v}_i \otimes \mathbf{a}_i,
	\label{eqs6}
	\end{align}
	where $\otimes$ represents the Kronecker product.

	\subsection{Design of the Codebook}
	\label{Sec1b}
	While the elements of the codebook $\mathbf{A}$ are selected in a similar fashion to \cite{pradhan2020polar}, unlike the codebook in \cite{pradhan2020polar} where the empirical variance of all the columns of $\mathbf{A}$ are normalized, we divide the columns of $\mathbf{A}$ into $m$ groups, and assign different power levels to each. There are $l_k$ columns with power level $P_k$ in the $k$th group, with $k=1,2,\dots,m$. Hence, with a uniform selection, the probability of choosing a column with power $P_k$ from the codebook is
	\begin{align}
	\mathrm{Pr}(\|\mathbf{a}_i\|^2=P_k)=\dfrac{l_k}{2^{B_s}}.
	\label{eqs2}
	\end{align}
	\indent Since $K_a$ active users pick columns of $\mathbf{A}$ randomly, the number of users with power $P_k$ can be approximated by
	\begin{align}
	K_k \approx \dfrac{l_k}{2^{B_s}}K_a,
	\label{eqs3}
	\end{align}
	\noindent for $K_a\gg1$. Optimization of the codebook parameters $l_k$ and $P_k$ is described in the next section.
	
	\subsection{Decoder}
	
	The decoder is composed of three parts. A newly proposed covariance-based detector identifies the set of spreading sequences employed by the active users; a minimum mean square error (MMSE) estimator is used to produce the soft estimates corresponding to the detected sequences; and finally, the estimated symbols are fed as input to the channel decoder, implemented as a polar list decoder. Successive interference cancellation (SIC) is employed to remove the contribution of the successfully decoded messages from the received signal at each step, and the procedure is repeated until messages of all the active users are decoded or there are no successfully decoded users in an iteration. 
	\subsubsection{Covariance-based Detector}
	As an alternative to the energy detector in \cite{pradhan2020polar}, which requires searching through possible codeword sequences, we propose an approach which generates an estimate of the covariance matrix of the received signal, and declares the signatures corresponding to the largest diagonal entries as the active ones. Namely, we write the remaining signal after last SIC step, $\mathbf{Y}^\prime \in \mathbb{R}^{n_s \times n_c}$, as
	\begin{align}
	\mathbf{Y}^\prime = \mathbf{A}_{a}\mathbf{V}+\mathbf{Z},\label{eq9}
	\end{align}
	\noindent where $\mathbf{A}_{a}\in \mathbb{R}^{n_s \times K}$ and $\mathbf{V}\in \mathbb\{\pm 1\}^{K \times n_c}$ are constructed by aggregating the signatures and codewords of the remaining $K$ active users; and, $\mathbf{Z}\in \mathbb{R}^{n_s \times n_c}$ is the noise matrix with independent and identically distributed zero-mean unit-variance Gaussian random variables as its elements. Note that since via a CRC check, we only remove the contribution from the correct decoded message at the SIC steps, \eqref{eq9} holds for all the iterations (with the remaining users' messages). We form
	\begin{align}
	\mathbf{C} = 	\mathbf{A}_{N }^T\mathbf{Y}^{\prime} \mathbf{Y}^{\prime T}		\mathbf{A}_{N }  ,
	\label{eqs10new}
	\end{align}
	\noindent where $\mathbf{A}_{N }$ is obtained by scaling the columns of the codebook to 1 (after removing signatures of correctly decoded users) to prevent higher detection probability for the sequences with greater power levels. The covariance-based detector outputs $K_{\gamma}=K_r+K_\delta$ signatures corresponding to largest diagonal elements of $\mathbf{C}$, where $K_r$ is the number of remaining users up to the current iteration and $K_\delta$ is a small integer.\\ \indent Let us comment on the complexity of the proposed approach. The number of multiplications required to calculate the diagonal entries of matrix $\mathbf{C}$ in \eqref{eqs10new} is $N_c = n_s^2n_c+2^{B_s}n_s^2+2^{B_s}n_s$, while the number of multiplications required in the energy detector \cite{pradhan2020polar} is $N_e = n_sn_c2^{(g+B_s)}$. Here $g$ denotes the length of each partition in the energy detector approach. Clearly, for moderate values of $g$, e.g., for $g= 8$, the complexity of the energy detector is significantly higher than that of the newly proposed one. For small $g$ values, however, the complexities of the two approaches are in the same order.
	\subsubsection{Channel Decoder and SIC}
	Accumulating the signatures declared by the covariance-based detector in the matrix $\hat{\mathbf{A}}_{\mathcal{D}}\in \mathbb{R}^{n_s \times (K_r+K_\delta)}$, the MMSE estimate of $\mathbf{V}$ is obtained by \cite{pradhan2020polar}
	\begin{align}
	\hat{\mathbf{V}}  = \hat{\mathbf{A}}_{\mathcal{D}}^T\hat{\mathbf{C}}_y^{-1}\mathbf{Y}^\prime , \label{eq12}
	\end{align}
	\noindent with $\hat{\mathbf{C}}_y=(\mathbf{I}_{n_s}+ \hat{\mathbf{A}}_{\mathcal{D}} \hat{\mathbf{A}}_{\mathcal{D}}^T)$,
	\noindent where the $i$th row of $\hat{\mathbf{V}}$ (denoted by $\hat{\mathbf{v}}_i$) is the estimated codeword of user $i$. We treat $\hat{\mathbf{v}}_i$ as the output of an AWGN channel with noise variance $\hat{\sigma}_i^2$, where $\hat{\sigma}_i^2$ is obtained by picking the $i$th diagonal entry of the mean square error (MSE) matrix, $\mathrm{MSE} = \mathbf{I}- \hat{\mathbf{A}}_{\mathcal{D}}^T \mathbf{C}_y^{-1}\hat{\mathbf{A}}_{\mathcal{D}}$. The $i$th user's message is then decoded by feeding $\mathbf{d}_i = 2\hat{\mathbf{v}}_i/\hat{\sigma}_i^2$ as the set of  log-likelihood ratios (LLRs) to the list decoder.\\
	\indent After polar decoding, the successfully decoded codewords that satisfy the CRC check are removed from the received signal. The residual received signal is then passed back to the covariance-based detector for the next iteration. This procedure is repeated until there are no successful decoding results or all the active users are decoded.\\
	\indent We finally note that there could be some false signatures declared by the covariance-based detector. In practice, during the decoding process, these falsely detected signatures with higher power levels can adversely affect the decoding performance of the users with lower power levels. To alleviate this effect, we run the decoder $m+1$ times with indices $i=0,1,...,m$. At the $i$th step, the signatures of the $i$ groups with the larger power levels are removed from the codebook when using the covariance-based detector. Therefore, the matrix $\hat{\mathbf{A}}_{\mathcal{D}}$ does not have the spreading sequences with high power levels in it. This helps the MMSE estimator in \eqref{eq12} to avoid possible negative effects of the falsely detected signatures with high power levels particularly when most of the users with these power levels are removed from the received signal by SIC. Note again that the falsely detected signatures do not negatively affect the SIC process since CRC check is used to verify the correctness of decoder outputs. The details of the decoding procedure are given as a pseudo-code in Algorithm I.
	
	\section{Design of Spreading Sequences: Optimal Power Allocation}

	\subsection{Selection of Optimal Parameters}
	Suppose that active users are divided into $m$ groups consisting of $K_1, K_2, ...,K_m$ users with power levels $P_1\leq P_2\leq ...\leq P_m$, where $K_1+K_2+...+K_m=K_a$. Note that during the interference cancellation procedure, the users wih the highest power levels are expected to be decoded first. Focusing on \eqref{eqs6}, we notice that $\mathbf{x}_i$ is obtained by normalizing a zero-mean Gaussian vector, hence we model the aggregate interference as a Gaussian random vector added to each user's signal. For instance, for the group with the highest power level, $K_m$ users with power $P_m$ experience interference with variance $\sigma^2_{m}=1+K_1P_1+K_2P_2+...+K_{m-1}P_{m-1}$. After successful decoding and  interference cancellation, we decode the $K_{m-1}$ users with power level $P_{m-1}$, which experience interference with variance $\sigma^2_{m-1}=1+K_1P_1+K_2P_2+...+K_{m-2}P_{m-2}$. Similarly, for the $j$th group, the $K_j$ users with power $P_j$ experience interference with variance $\sigma_j^2 = 1+\sum_{i=1}^{j-1}K_iP_i$. From \eqref{eqs10} in Appendix \ref{appendixA}, the minimum required power for the $j$th group becomes
	\begin{align}
	P_j =\dfrac{{ \alpha_{\mathrm{min}}(K_j)}}{1-(K_j-1)\alpha_{\mathrm{min}}(K_j)}\left(1+\sum_{i=1}^{j-1}K_iP_i\right),
	\end{align}
	\noindent which can be rewritten as
	\begin{align}
	P_j &=\gamma_j\prod_{i=1}^{j-1}(1+K_i\gamma_i) \ \forall j = 1, 2,\hdots,m   ,
	\end{align}
	\noindent where $\gamma_j = \dfrac{{ \alpha_{\mathrm{min}}(K_j)}}{1-(K_j-1)\alpha_{\mathrm{min}}(K_j)}$, and $\alpha_{\mathrm{min }}(K_j)$ is the minimum required signal-to-interference-plus-noise ratio (SINR) for achieving a target PUPE in a group with $K_j$ users. Note that $\alpha_{\mathrm{min}}(K_j)$ depends on the particular coding and transmission scheme employed, and can be determined via simulations. Therefore, the total power can be written as a function of $K_1, K_2, ..., K_m$ as
	\begin{align}
	\nonumber
	P_T &= K_1P_1+K_2P_2+...+K_mP_m\\
	&=\prod_{i=1}^m (1+K_i\gamma_i)-1.\label{eqs88}
	\end{align}
	\indent We can find values of $K_1, K_2, ... , K_m$ which minimize the total power by solving the following  optimization problem:
	\begin{align}
	(\hat{K}_1, ..., \hat{K}_m) = \argmin_{K_1, ..., K_m}P_T,\ \  \mathrm{s.t.} \ \sum_{i=1}^m K_i=K_a.\label{eqs12}
	\end{align}
	\indent To proceed further, we relax the integer constraint on $K_j$'s, and use the method of Lagrange multipliers. The cost function to be minimized is
	\begin{align}
	\mathbf{J}(K_1,\hdots,K_m) = \prod_{i=1}^m (1+K_i\gamma_i)+\lambda \left( \sum_{i=1}^m K_i-K_a\right)\label{eqs14}
	\end{align}
	\noindent where $\lambda$ is the Lagrange multiplier. Setting the derivative of the Lagrangian function to zero gives
	\begin{align}
	\dfrac{K_1\gamma^\prime_1+\gamma_1}{1+K_1\gamma_1}=\dfrac{K_2\gamma^\prime_2+\gamma_2}{1+K_2\gamma_2}=...=\dfrac{K_m\gamma^\prime_m+\gamma_m}{1+K_m\gamma_m}=-\dfrac{\lambda}{P_T+1}.
	\label{eqs11}
	\end{align}
	\indent Strictly speaking, while $\alpha_{\mathrm{min}}(K_j)$ is a function of $K_j$, this value is almost a constant (for a given target PUPE level). Dropping this dependence, after some calculations, we obtain
	\begin{align}
	\dfrac{K_i\gamma^\prime_i+\gamma_i}{1+K_i\gamma_i}=\dfrac{\alpha_{\mathrm{min}}}{1-(K_i-1)\alpha_{\mathrm{min}}}.
	\end{align}
	
	\indent Since $\dfrac{\alpha_{\mathrm{min}}}{1-(K_i-1)\alpha_{\mathrm{min}}}$ is a one-to-one function of $K_i$, the optimal values of $K_i$ satisfy	$
	K_1=K_2=...=K_m=K_0,
	$ with $K_0 = K_a/m$. We can also argue that (with a constant $\alpha_{\mathrm{min }}$), the Hessian for $\mathbf{J}(K_1,\hdots,K_m)$ is given as
	\begin{align}
	\nabla^2 (\mathbf{J})=(P_T+1) \mathrm{diag}\left(\dfrac{K_1{\gamma}_1''+2{\gamma}_1'}{1+K_1\gamma_1},\hdots ,\dfrac{K_m{\gamma}_m''+2{\gamma}_m'}{1+K_m\gamma_m}\right)\label{eqs89}
	\end{align}
	with 
	\begin{align}
	{\gamma}_i' &= \dfrac{\partial \gamma_i}{\partial K_i}=\left(\dfrac{\alpha_{\mathrm{min}}}{1-(K_i-1)\alpha_{\mathrm{min}}}\right)^2, \\
	{\gamma}_i'' &=\dfrac{\partial^2 \gamma_i}{\partial K_i^2}=2\left(\dfrac{\alpha_{\mathrm{min}}}{1-(K_i-1)\alpha_{\mathrm{min}}}\right)^3, 
	\end{align}
	\noindent where $\mathrm{diag}(\zeta_1, ...,\zeta_l)$ is an $l\times l$ diagonal matrix with the $i$th diagonal element $\zeta_i$. From \eqref{eqs10}, we observe that $\dfrac{\alpha_{\mathrm{min}}}{1-(K_i-1)\alpha_{\mathrm{min}}}$ is proportional to the minimum required power, which is always positive. Thus, the Hessian in \eqref{eqs89} is a positive definite matrix, and \eqref{eqs12} is verified to be a convex optimization problem. \\
	\indent Determining that the number of users in all the groups should be the same, we optimize the number of groups via 
	\begin{align}
	m =\min_{\hat{m}} \left(1+\dfrac{K_a\gamma_0}{\hat{m}}\right)^{\hat{m}}.
	\label{eqs13}
	\end{align}
	\vspace{-10.0 mm}
	\subsection{General case}
	
	We highlight that, even though the optimal PA approach is applied to the random spreading scheme, the basic ideas can be extended to other unsourced MAC scenarios (for which the interference is treated as noise) as well by solving \eqref{eqs13} for the specific $\alpha_{\mathrm{min}}(K)$ values.
	\vspace{-2.0 mm}
	\section{Numerical Results}
\begin{figure}[t!]
		\centering
		\includegraphics[width=.975\linewidth]{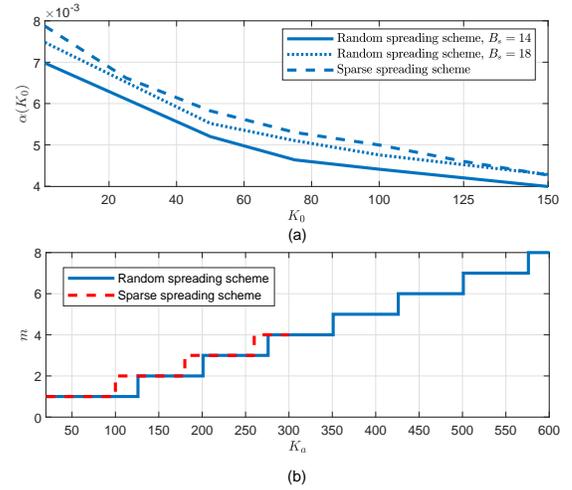}
		\caption{{\small (a) The required SINR and (b) optimal number of groups for random spreading scheme with $n_s = 117$ and $n_c = 256$ and sparse spreading scheme in \cite{zheng2020polar}.}}
		\label{Fig1}
	\end{figure}

	\vspace{-1.0 mm}
	In this section, we study the performance of the proposed approach and compare it with other existing solutions for unsourced MAC. The total number of channel uses is selected as $n \cong 30000$ ($n_s = 117$ and $n_c = 256$), the number of active users is $150\leq K_a\leq 600$, the list size for the polar decoder is $L=512$, and the target PUPE is set to $P_e = 0.05$. The number of information bits per message is chosen as $B=100$, where $B_s$ is selected as $B_s = 14$ for $150\leq K_a < 200$, $B_s = 15$ for $200\leq K_a \leq 250$, $B_s = 16$ for $250< K_a \leq 350$, $B_s = 17$ for $350< K_a \leq 500$, and $B_s = 18$ for $500< K_a \leq 600$. Note that, the probability that three or more users are in collision is very small for the selected values of $B_s$, hence their effects can be neglected. Also, as discussed in \cite{pradhan2020polar}, the random spreading scheme with channel coding rate lower than $1/2$ is able to resolve the possible collision of two users, providing a good performance.\\		
	\indent To obtain $\alpha_{\mathrm{min}}(K_0)$ for the random spreading scheme, the method in \cite{pradhan2020polar} is run for different values of $K_0$, and the result is shown in Fig.~\ref{Fig1}(a) for $B_s = 14$ and $18$. One should keep in mind that the required SINR must be computed for different values of $B_s$ separately, because in the random spreading scheme, it is sensitive to $B_s$. Plugging $\alpha_{\mathrm{min}}(K_0)$ into \eqref{eqs13}, the optimal number of groups, $m$, is obtained (as depicted in Fig. \ref{Fig1}(b)). Clearly, for $K_a\leq 125$, only one group should be employed, however, beyond that number of users, employing more than one group with different power levels becomes optimal. 
	
\begin{figure}[t!]
		\centering
		\includegraphics[width=.975\linewidth]{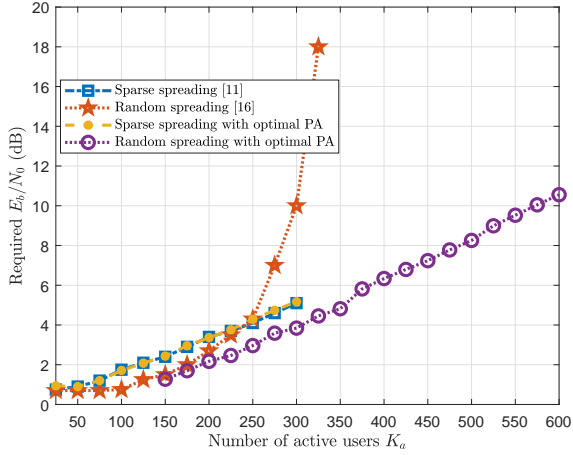}
		\caption{{\small The required $E_b/N_0$ as a function of the number of active users in the sparse spreading scheme \cite{zheng2020polar}, the random spreading scheme \cite{pradhan2020polar}, the sparse spreading with optimal PA, and the random spreading with optimal PA.}}
		\label{Fig3}
	\end{figure}

	We also apply our approach to the sparse spreading idea of~\cite{zheng2020polar} to determine the optimal number of groups and the corresponding power levels without any need for extensive simulations. The results are depicted in Fig.~\ref{Fig1}. We compare the required $E_b/N_0$ of the scheme in \cite{zheng2020polar}, random spreading scheme in \cite{pradhan2020polar}, sparse spreading with optimal PA, and random spreading with optimal PA in Fig.~\ref{Fig3}. It can be inferred from this result that applying optimal PA on the random spreading scheme decreases the required $E_b/N_0$ of the system remarkably, especially, for the larger number of active users, i.e., for $K_a>225$. Moreover, it is shown that the required $E_b/N_0$ is the same for the sparse spreading scheme with optimal PA and sparse spreading scheme in \cite{zheng2020polar} where the parameters are empirically chosen. 
	
\begin{figure}[t!]
		\centering
		\includegraphics[width=.975\linewidth]{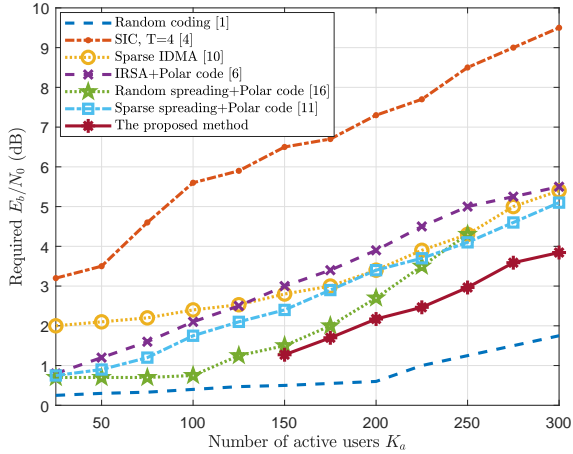}
		\caption{{\small The required $E_b/N_0$ as a function of the number of active users for the proposed scheme and other solutions.}}
		\label{Fig4}
	\end{figure}

	In Fig.~\ref{Fig4}, the performance of the proposed random spreading solution with optimal PA is compared with the other existing unsourced MAC schemes. The result clearly indicates that the proposed method offering superior performance, particularly, when the number of active users is large. \\
	\indent We further note our observation that the detection probability of the covariance-based detector is slightly higher than that of the energy detector for manageable values of $g$, and for the specific parameters of $n_s$, $n_c$, and $B_s$ in this section. Regarding its complexity, for $K_a = 500$ and $B_s = 17$, the total number of multiplications required by the proposed covariance-based detector and the energy detector with $g=1$ are $N_c=1.81\times 10^9$ and $N_e=7.85\times 10^9$, respectively. That is, the computational complexity of the covariance-based detector and energy detector with $g=1$ are comparable.	
	
	\begin{algorithm}
		\begin{small}
			\caption{{\small Pseudo-code for the decoder steps.}} 
			\textbf{Input:} $\mathbf{Y}$, $\mathbf{A}$, and $K_a$.\\	
			$\mathrm{flag} = 1$.\\	
			$K_r = K_a$.\\
			$\mathbf{Y}^\prime=\mathbf{Y}.$ \\
			\While( {Decoding}){$\mathrm{flag}=1$}
			{
				$K_x = 0$.\\
				\For( ){{$i = 0,1, \hdots, m $}}
				{
					\begin{itemize}
						\item $i$ groups with higher powers are removed from $\mathbf{A}_{N }$.
						\item $\mathcal{S}\  = \emptyset$.
						\item $\hat{\mathbf{A}}_{\mathcal{D}}$ is the output of covariance-based detector.
						\item Calculate $\hat{\mathbf{V}} = \hat{\mathbf{A}}_{\mathcal{D}}^T\hat{\mathbf{C}}_y^{-1}\mathbf{Y}^\prime $.
						\item Feed $\mathbf{d}_i = 2\hat{\mathbf{v}}_i/\hat{\sigma}_i^2$ to the list decoder.
						\item update $\mathcal{S}$ as the set of decoded codewords.
						
						\item $K_r = K_r-|\mathcal{S}|$ and $K_x = K_x+|\mathcal{S}|$.
						\item $\mathbf{Y}^\prime = \mathbf{Y}^\prime-\mathbf{A}_{\mathcal{S}}\mathbf{V}_{\mathcal{S}}$.
					\end{itemize}
					\If{$K_x\geq 1$ }{ break.}
				}
				\If{$K_x=0$ $\mathrm{or}$ $K_r=0$}{$\mathrm{flag}=0$.}
			}
		\end{small}
	\end{algorithm}
	\vspace{-2.0 mm}
	\section{Conclusions}
	\vspace{-1.0 mm}
	We have studied unsourced MAC with polar codes and random spreading, and developed an approach which divides the active users into different groups with varying transmit power levels. The optimal number of groups and the power levels are selected through a suitably formulated optimization problem, and it is shown by numerical evaluations that the idea of using different power levels provides a significant reduction in the required $E_b/N_0$, particularly, for systems with a large number of active users.\\
	\vspace{-5.8mm}
	\begin{appendices}
		\section{Obtaining Minimum Required Power in a group}
		\label{appendixA}
		Since we are treating interference as noise, from the perspective of a user, the SINR becomes the important performance metric. In a group with $K_j$ users with power level $P_j$, and noise variance $\sigma^2_j$, we need
		\vspace{-1.8mm}
		\begin{align}
		\alpha_{\mathrm{min}}(K_j) \leq \dfrac{P_j}{\sigma^2_j+{(K_j-1)}P_j}.
		\label{eqs8}
		\end{align}
		\indent Since our objective is to minimize the total power, we can select the power level for each user as 
		\vspace{-1.8mm}
		\begin{align}
		P_j =\dfrac{{  \alpha_{\mathrm{min}}(K_j)}}{1-(K_j-1)\alpha_{\mathrm{min}}(K_j)}\sigma^2_j.
		\label{eqs10}
		\end{align}
	\end{appendices}
	\vspace{-7.8mm}
	\bibliographystyle{IEEEtran}
	\bibliography{ahmadi}

\begin{thebibliography}{10}
\providecommand{\url}[1]{#1}
\csname url@samestyle\endcsname
\providecommand{\newblock}{\relax}
\providecommand{\bibinfo}[2]{#2}
\providecommand{\BIBentrySTDinterwordspacing}{\spaceskip=0pt\relax}
\providecommand{\BIBentryALTinterwordstretchfactor}{4}
\providecommand{\BIBentryALTinterwordspacing}{\spaceskip=\fontdimen2\font plus
\BIBentryALTinterwordstretchfactor\fontdimen3\font minus
\fontdimen4\font\relax}
\providecommand{\BIBforeignlanguage}[2]{{%
\expandafter\ifx\csname l@#1\endcsname\relax
\typeout{** WARNING: IEEEtran.bst: No hyphenation pattern has been}%
\typeout{** loaded for the language `#1'. Using the pattern for}%
\typeout{** the default language instead.}%
\else
\language=\csname l@#1\endcsname
\fi
#2}}
\providecommand{\BIBdecl}{\relax}
\BIBdecl

\bibitem{polyanskiy2017perspective}
Y.~Polyanskiy, ``A perspective on massive random-access,'' in \emph{Proc. IEEE
ISIT}, pp. 2523--2527, Aachen, Germany, 2017.

\bibitem{calderbank2020chirrup}
R.~Calderbank and A.~Thompson, ``Chirrup: a practical algorithm for unsourced
multiple access,'' \emph{Information and Inference: A Journal of the IMA},
vol.~9, no.~4, pp. 875--897, Dec. 2020.

\bibitem{vem2017user}
A.~Vem, K.~R. Narayanan, J.~Cheng, and J.-F. Chamberland, ``A user-independent
serial interference cancellation based coding scheme for the unsourced random
access {G}aussian channel,'' in \emph{Proc. IEEE ITW}, pp. 121--125, Kaohsiung, Taiwan, 2017.

\bibitem{vem2019user}
A.~Vem, K.~R. Narayanan, J.-F. Chamberland, and J.~Cheng, ``A user-independent
successive interference cancellation based coding scheme for the unsourced
random access {G}aussian channel,'' \emph{IEEE Transactions on
Communications}, vol.~67, no.~12, pp. 8258--8272, Dec. 2019.

\bibitem{ordentlich2017low}
O.~Ordentlich and Y.~Polyanskiy, ``Low complexity schemes for the random access
{G}aussian channel,'' in \emph{Proc. IEEE
ISIT}, pp. 2528--2532, Aachen, Germany, 2017.

\bibitem{marshakov2019polar}
E.~Marshakov, G.~Balitskiy, K.~Andreev, and A.~Frolov, ``A polar code based
unsourced random access for the {G}aussian MAC,'' in \emph{Proc. IEEE VTC}, pp. 1--5, Honolulu, USA, 2019.

\bibitem{amalladinne2018coupled}
V.~K. Amalladinne, A.~Vem, D.~K. Soma, K.~R. Narayanan, and J.-F. Chamberland,
``A coupled compressive sensing scheme for unsourced multiple access,'' in
\emph{Proc. IEEE ICASSP}, pp. 6628--6632, Calgary, Canada, 2018.

\bibitem{amalladinne2018coded}
V.~K. Amalladinne, J.-F. Chamberland, and K.~R. Narayanan, ``A coded compressed
sensing scheme for uncoordinated multiple access,'' \emph{arXiv preprint
arXiv:1809.04745}, 2018.

\bibitem{9432925}
A.~Fengler, P.~Jung, and G.~Caire, ``Sparcs for unsourced random access,''
\emph{IEEE Transactions on Information Theory}, 2021.

\bibitem{pradhan2019joint}
A.~Pradhan, V.~Amalladinne, A.~Vem, K.~R. Narayanan, and J.-F. Chamberland, ``A joint graph based coding scheme for the unsourced random access {G}aussian channel,'' in \emph{Proc. IEEE GLOBECOM}, pp. 1--6, Waikoloa, USA, 2019.
\bibitem{zheng2020polar}
M.~Zheng, Y.~Wu, and W.~Zhang, ``Polar coding and sparse spreading for massive
unsourced random access,'' in \emph{Proc. IEEE VTC}, pp. 1--5, Victoria, Canada, 2020.
\bibitem{truhachev2020low}
D.~Truhachev, M.~Bashir, A.~Karami, and E.~Nassaji, ``Low-complexity coding and
spreading for unsourced random access,'' \emph{IEEE Communications Letters}, vol.~25, no.~3, pp. 774--778, Mar. 2021.

\bibitem{tanc2021massive}
A.~K. Tanc and T.~M. Duman, ``Massive random access with trellis based codes and random signatures,'' \emph{IEEE Communications Letters}, vol.~25, no.~5, pp. 1496--1499, May. 2021.
\bibitem{han2021sparse}
Z.~Han, X.~Yuan, C.~Xu, S.~Jiang, and X.~Wang, ``Sparse kronecker-product
coding for unsourced multiple access,'' \emph{arXiv preprint
arXiv:2103.04722}, 2021.
\bibitem{ebert2021stochastic}
J.~R. Ebert, V.~K. Amalladinne, S.~Rini, J.-F. Chamberland, and K.~R.
Narayanan, ``Stochastic binning and coded demixing for unsourced random
access,'' \emph{arXiv preprint arXiv:2104.05686}, 2021. 
\bibitem{pradhan2020polar}
A.~K. Pradhan, V.~K. Amalladinne, K.~R. Narayanan, and J.-F. Chamberland,
``Polar coding and random spreading for unsourced multiple access,'' in
\emph{Proc. IEEE ICC}, pp. 1--6, Dublin, Ireland, 2020.
\bibitem{pradhan2021ldpc}
A.~Pradhan, V.~Amalladinne, K.~R. Narayanan, and J.-F. Chamberland, ``LDPC codes with soft interference cancellation for uncoordinated
unsourced multiple access,'' \emph{arXiv preprint arXiv:2105.13985}, 2021.

\end{thebibliography}
\end{document}